\documentstyle[fleqn,amssymb,amsbsy]{article}

\newcommand{\be}{\begin{equation}}
\newcommand{\ee}{\end{equation}}
\newcommand{\ba}{\begin{array}}
\newcommand{\ea}{\end{array}}
\newcommand{\bea}{\begin{eqnarray}}
\newcommand{\eea}{\end{eqnarray}}
\newcommand{\bean}{\begin{eqnarray*}}
\newcommand{\eean}{\end{eqnarray*}}
\newcommand{\lp}{\left(}
\newcommand{\rp}{\right)}
\newcommand{\ls}{\left[}
\newcommand{\rs}{\right]}
\newcommand{\lc}{\left\{}
\newcommand{\rc}{\right\}}

\newcommand{\la}{\langle}
\newcommand{\La}{\left\la}
\newcommand{\ra}{\rangle}
\newcommand{\Ra}{\right\ra}
\renewcommand{\d}{{\rm d}}
\newcommand{\e}{{\rm e}}

\newcommand{\intl}{\int\limits}

\newcommand{\Imm}{\,\Im{\frak m}\,}

\newcommand{\im}{{\rm i}}

\newcommand{\veps}{\varepsilon}

\newcommand{\vrho}{\varrho}
\newcommand{\mGamma}{{\mit\Gamma}}

\newcommand{\mTheta}{{\mit\Theta}}

\newcommand{\mPhi}{{\mit\Phi}}

\newcommand{\Chi}{\mbox{\parbox[c]{0.8em}{\large$\chi$}}}
\newcommand{\const}{\mathop{{\rm const}}\nolimits}
\newcommand{\ddt}{\frac{\partial}{\partial t}}
\newcommand{\dd}[1]{\frac{\partial}{\partial #1}}
\newcommand{\refp}[1]{(\ref{#1})}
\newcommand{\ds}{\displaystyle}
\newcommand{\scs}{\scriptstyle}
\newcommand{\bmv}[1]{{\boldsymbol #1}}
\newcommand{\xnt}{\lp x^N;t\rp}

\newcommand{\El}{{\cal L}}

\title{\bf The modified group expansions for construction of solutions to the
BBGKY hierarchy}

\author{A.E.Kobryn, I.P.Omelyan, M.V.Tokarchuk\\[1ex]\normalsize
\it Institute for Condensed Matter Physics of the Ukrainian National Academy
of Sciences\\ \normalsize\it 1~Svientsitskii St., UA--290011 Lviv--11,
Ukraine}

\date{\today}

\begin{document}

\maketitle

\begin{abstract}
A solution to the BBGKY hierarchy for nonequilibrium distribution functions
is obtained within modified boundary conditions. The boundary conditions
take into account explicitly both the non\-equ\-i\-li\-bri\-um one-particle
distribution function as well as local conservation laws. As a result,
modified group expansions are proposed. On the basis of these expansions,
a generalized kinetic equation for hard spheres and a generalized
Bogolubov-Lenard-Balescu kinetic equation for a dense electron gas are
derived within the polarization approximation.


\end{abstract}

\section{Introduction}
\setcounter{equation}{0}

The construction of kinetic equations for {\em dense} gases, liquids and
plasma, where a small parameter (density or interaction) is absent, still
remains one of the most significant complication in the kinetic theory of
classical systems. A number of difficult tasks centre around it. The study
of solutions to the BBGKY hierarchy for nonequilibrium distribution functions
in view of interparticle correlations could be considered as a way to solve
this problem.

Recently, a kinetic equation of the revised Enskog theory for a dense
system of hard spheres and an Enskog-Landau kinetic equation for a dense
system of charged hard spheres have been obtained from the BBGKY hierarchy
in the ``pair'' collisions approximation \cite{1,2,2a}. It is necessary to
notice that this approximation does not correspond to the usual two-particle
approximation, inherent in the Boltzmann theory, because an essential part
of the many-particle correlations is implicitly taken into account by the 
pair quasiequilibrium distribution function $g_2(\bmv{r}_1,\bmv{r}_2;t)$.

To analyze solutions to the BBGKY hierarchy \cite{1,2} in higher 
approximations
on interparticle correlations, it is more convenient to use the concept of
group expansions \cite{3,4,5}. This has been applied to the BBGKY hierarchy
in previous investigations \cite{3,4,5,6,7,8,9,10,11,12,13,14,15,16} using
boundary conditions which correspond to the weakening correlations principle
by Bogolubov \cite{17}. The same conception has been involved in papers
of Zubarev and Novikov \cite{4,18,19,20}, where a diagram method for 
obtaining solutions to the BBGKY hierarchy was developed. Moreover,
in papers by D.N.Zubarev et al. \cite{21,22} a {\em consistent}
description of kinetics and hydrodynamics has been proposed. It is based on
a new formulation of boundary conditions for the Liouville equation and 
BBGKY hierarchy. In particular, this formulation takes into account
interparticle correlations which are connected with local conservation
laws.

In this paper, the conception of group expansions will be applied to the
BBGKY hierarchy with modified boundary conditions which take into account
both the nonequilibrium behaviour of one-particle distribution function
and local conservation laws (sec. 2). In sections 3 and 4 it will be
demonstrated on the basis of the modified group expansions how to obtain
a generalization of some known kinetic equations for hard spheres and
homogeneous plasma to the case of high densities. Advantages and
shortcomings of the proposed approach are discussed concisely in conclusion 
and at the end of some sections.

\section{The BBGKY hierarchy with modified boundary conditions}
\setcounter{equation}{0}

The BBGKY hierarchy of equations for nonequilibrium distribution functions
of classical interacting particles has been obtained in the paper \cite{21}
on the basis of assembling of time retarded solutions for the Liouville
equation with the modified Bogolubov's condition of weakening correlations
between particles. According to the Zubarev's nonequilibrium statistical
operator method \cite{z,zz}, the total nonequilibrium distribution function
$\vrho\lp x^N;t\rp$ for all $N$ particles of a system satisfies the
following asymptotic condition:
%
%
\be
\lim_{t_0\to-\infty}\exp\left(\im L_Nt_0\right)
        \left(\vrho\left(x^N,t_0\right)-\vrho_q\left(x^N,t_0\right)\right)=0.
        \label{bc}
\ee
Here, $\im=\sqrt{-1}$, and the limit $t_0\to-\infty$ is taken after 
thermodynamic limiting transition $N\to\infty$, $V\to\infty$, 
$N/V\to\const$; $L_N$ is the Liouville operator:
%
%
\be
L_N=\sum\limits_{j=1}^NL(j)+
        \frac 12\mathop{\sum_{j=1}^N\sum_{k=1}^N}\limits_{j\ne k}L(j,k),
        \;\; L(j)=-\im\frac{\bmv{p}_j}{2m}\dd{\bmv{r}_j},\;\;
        L(j,k)=\im\frac{\partial\mPhi\left(|\bmv{r}_{jk}|\right)}
        {\partial\bmv{r}_{jk}}\left(\dd{\bmv{p}_j}-\dd{\bmv{p}_k}\right),
        \label{lo}
\ee
${\mPhi}_{jk}$ is the interaction energy between two particles $j$ and $k$;
$x_j=\{\bmv{r},\bmv{p}\}$ is the set of phase variables (coordinates and
momenta).

Let us consider time scales when details of initial state
$\vrho\left(x^N,t_0\right)$ become not so important, i.e., when $t \gg t_0$.
Then, to avoid the dependency on $t_0$, let us average the
formal solution to the Liouville equation $\vrho\lp x^N;t\rp=
\e^{-\im L(t-t_0)}\vrho\lp x^N;t_0\rp$ with respect to initial times
in the range $t_0\div t$ and make the limiting transition $t_0-t\to\infty$
\cite{z,zz},
%
%
\be
\vrho\xnt=\veps\intl_{-\infty}^0\d t'\;\e^{\ds\veps t'}\e^{\ds\im L_Nt'}
        \vrho_{\rm q}\lp x^N;t+t'\rp,\label{e2.6}
\ee
where $\veps$ tends to $+0$ after the thermodynamic limiting transition.
The quasiequilibrium distribution function $\vrho_{\rm q}\lp x^N;t\rp$ is
determined from the condition of maximum for informational entropy at fixed
values of the single particle distribution function $f_1\lp x_1;t\rp$ and
average density of the interaction energy
$\la{\cal E}_{\rm int}(\bmv{r})\ra^t$,
$\la\ldots\ra^t=\int\d\mGamma_N\ldots\vrho\lp x^N;t\rp$,
$\d\mGamma_N=(\d x)^N/N!$. This corresponds to taking into account
correlations, related to the conservations laws of hydrodynamical variables
for particles density $n(\bmv{r};t)$, momentum $\bmv{j}(\bmv{r};t)$ and full
energy ${\cal E}(\bmv{r};t)$ \cite{22}.

The boundary condition (\ref{bc}) is equivalent to
the transition from the Liouville equation to a modified one \cite{z,zz}:
%
%
\be
\lp\ddt+\im L_N\rp\vrho\lp x^N,t\rp=-\varepsilon\Big(\vrho\lp x^N,t\rp-
        \vrho_{\rm q}\lp x^N,t\rp\Big).
        \label{mle}
\ee
This equation contains a small source on the right-hand side, which
destroys the invariance with respect to time inversion.

The choice of $\vrho_{\rm q}\xnt$ depends mainly on nonequilibrium state
of the system under consideration. In the case of low density gases, where
times of free motion are essentially greater then collision times,
higher-order distribution functions of particles become depend on
time via oneparticle distribution functions only \cite{4}. It does mean that
an abbreviated description of nonequilibrium states is available, and
the total nonequilibrium distribution function depends on time via
$f_1(x;t)$. In such a case, the quasiequilibrium distribution function
$\vrho_{\rm q}\xnt$ reads \cite{4}:
%
%
\be
\vrho_{\rm q}\xnt=\prod_{j=1}^N\frac{f_1(x_j;t)}{\e},\label{e2.8}
\ee
where $\e$ is the natural logarithm base. Then,
the Liouville equation with a small source \refp{mle} in view of
\refp{e2.8} corresponds to the abbreviated description of time
evolution of the system on a kinetic stage when only the oneparticle
distribution function is considered as a slow variable. However, there are
always additional quantities, which vary in time slowly, because they are
locally conserved.
In the case of an onecomponent system, the mass density $\rho(\bmv{r};t)$,
momentum $\bmv{j}(\bmv{r};t)$ and total energy ${\cal E}(\bmv{r};t)$ belong
to those quantities. At long times they satisfy the generalized hydrodynamics
equations. Generally speaking, the equation for $f_1(x;t)$ should be
conjugated with these equations. For low density gases, such a conjugation
can be made, 
in principle, with arbitrary precision in each order over density.
In high density gases and liquids, when a small parameter is absent, the
correlation times corresponding to hydrodynamic quantities become to be
commensurable with characteristic times of varying of oneparticle 
distribution 
functions. Therefore, in dense gases and liquids the kinetics and
hydrodynamics are closely connected between themselves and they
should be considered simultaneously. That is why, manyparticle
correlations, related to the local conservation laws of mass, momentum
and total energy, can not be neglected \cite{z,zz}. 
The local conservation laws
affect on kinetic processes, because of the interaction of selected
particles group with other particles of a system. This interaction is
especially important in the case of high densities, and it must be included
into consideration. Then at the construction of kinetic equations for
high densities, it is necessary to choose the abbreviated description of
nonequilibrium system in such a form to satisfy the true dynamics of
conserved quantities automatically. To this end, the
densities of hydrodynamic variables should be included together with the
oneparticle distribution function $f_1(x;t)$ into the set of parameters of 
the abbreviated description initially \cite{1,21,22}. 
The next phase functions correspond
to densities of hydrodynamic variables $\rho(\bmv{r};t)$,
$\bmv{j}(\bmv{r};t)$ and ${\cal E}(\bmv{r};t)$:
%
%
\be
\begin{array}{rcl}
\hat{\rho}(\bmv{r})&=&{\scs\int}\d\bmv{p}\;m\hat{n}_1(x),\qquad
\hat{\cal E}(\bmv{r})={\scs\int}\d\bmv{p}\;\frac{p^2}{2m}\hat{n}_1(x)
        +1/2{\scs\int}\d\bmv{r}'\;\d\bmv{p}\;\d\bmv{p}'\;
        \mPhi(|\bmv{r}-\bmv{r}'|)\hat{n}_2(x,x'),\\
\hat{\bmv{j}}(\bmv{r})&=&{\scs\int}\d\bmv{p}\;\bmv{p}\hat{n}_1(x).
\end{array}
\label{e2.9}
\ee
where $\hat{n}_1(x)$ and $\hat{n}_2(x,x')$ are the one- and two-particle
microscopic phase densities by Klimontovich \cite{12}. 
Relations \refp{e2.9} show
a distinctive role of potential interaction energy. In contrary to
$\rho(\bmv{r};t)=\la\hat{\rho}(\bmv{r})\ra^t$ and
$\bmv{j}(\bmv{r};t)=\la\hat{\bmv{j}}(\bmv{r})\ra^t$, nonequilibrium values
of the total energy ${\cal E}(\bmv{r};t)=\la\hat{\cal E}(\bmv{r})\ra^t$
can not be expressed via oneparticle distribution function
$f_1(x;t)=\la\hat{n}_1(x)\ra^t$ exclusively, because to evaluate
a potential part of ${\cal E}_{\rm int}(\bmv{r};t)=\la\hat{\cal
E}_{\rm int}(\bmv{r})\ra^t$ it is necessary to involve
the twoparticle
distribution function $f_2(x,x';t)$ $=\la\hat{n}_2(x,x')\ra^t$. Here
%
%
\be
\hat{\cal E}_{\rm int}(\bmv{r})=\frac 12
        \int\d\bmv{r}'\;\d\bmv{p}\;\d\bmv{p}'\;
        \mPhi(|\bmv{r}-\bmv{r}'|)\hat{n}_2(x,x')\label{e2.11}
\ee
is the density of potential energy of interaction. The next conclusion
can be formulated as follows. If oneparticle distribution function
$f_1(x;t)$ is chosen as a parameter of the abbreviated description, then
the density of interaction energy (\ref{e2.11}) can be considered as
an additional independent parameter.
One can find the quasiequilibrium distribution function
$\vrho_{\rm q}\xnt$ from the condition of extremum for the functional of
informational entropy $S_{\rm inf}(t)=
-\int\d\mGamma_N\;\vrho\xnt\ln\vrho\xnt$ at fixed average values of
$\la\hat{n}_1(x)\ra^t=f_1(x;t)$, $\la\hat{\cal E}_{\rm int}(\bmv{r})\ra^t=
{\cal E}_{\rm int}(\bmv{r};t)$ including the normalization condition
for $\vrho_{\rm q}\xnt$. After simple transformations the desired relation 
for the quasiequilibrium distribution function reads
%
%
\bea
\vrho_{\rm q}\xnt&=&\exp\lc-\Phi(t)-\int\d\bmv{r}\;\beta(\bmv{r};t)
        \hat{\cal E}_{\rm int}(\bmv{r})-\int\d x\;a(x;t)\hat{n}_1(x)\rc,
        \label{e2.13}\\
\Phi(t)&=&\ln\int\d\mGamma_N\;\exp\lc-\int\d\bmv{r}\;\beta(\bmv{r};t)
        \hat{\cal E}_{\rm int}(\bmv{r})-\int\d x\;a(x;t)\hat{n}_1(x)\rc,
        \label{e2.14}
\eea
where $\Phi(t)$, $\beta(\bmv{r};t)$, $a(x;t)$ are the Lagrange multipliers.
$\Phi(t)$ is the Massieu-Planck functional, which is determined from the
condition of normalization for $\vrho_{\rm q}\xnt$:
$\int\d\mGamma_N\vrho_{\rm q}\xnt=1$. Relation \refp{e2.13} for
$\vrho_{\rm q}\xnt$ has been obtained for the first time in \cite{21}. To
determine the physical meaning of parameters $\beta(\bmv{r};t)$ and $a(x;t)$
let us rewrite $\vrho_{\rm q}\xnt$ (\ref{e2.13}) in the form
%
%
\be
\vrho_{\rm q}\xnt=\exp\lc-\Phi(t)-\int\d\bmv{r}\;\beta(\bmv{r};t)
        \hat{\cal E}'(\bmv{r})-\int\d x\;a'(x;t)\hat{n}_1(x)\rc,
        \label{e2.16}
\ee
where $\hat{\cal E}'(\bmv{r})$ is the density of total energy in a reference
frame, which runs together a system element with the mass velocity
$\bmv{V}(\bmv{r};t)$ \cite{22,z}
%
%
\be
\hat{\cal E}'(\bmv{r})=\hat{\cal E}(\bmv{r})-
        \bmv{V}(\bmv{r};t)\hat{\bmv{j}}(\bmv{r})+\frac m2V^2(\bmv{r};t)
        \hat{n}(\bmv{r}).\label{e2.17}
\ee
Here $\hat{n}(\bmv{r})=\int\d\bmv{p}\;\hat{n}_1(x)$ is the density of
particles number. Parameters $\beta(\bmv{r};t)$ and $a'(x;t)$ in
(\ref{e2.16}) are defined from conditions of self-consistency, namely,
the equality of quasiaveraging values $\la\hat{n}_1(x)\ra^t_{\rm q}$ and
$\la\hat{\cal E}'(\bmv{r})\ra^t_{\rm q}$ to their real average values
$\la\hat{n}_1(x)\ra^t$, $\la\hat{\cal E}'(\bmv{r})\ra^t$:
%
%
\be
\begin{array}{rcl}
\la\hat{n}_1(x)\ra^t_{\rm q}&=&\la\hat{n}_1(x)\ra^t=f_1(x;t),\\
\la\hat{\cal E}'_1(\bmv{r})\ra^t_{\rm q}&=&
        \la\hat{\cal E}'_1(\bmv{r})\ra^t.\\
\end{array}
\label{e2.18}
\ee
Here $\la\ldots\ra^t_{\rm q}=\int\d\mGamma_N\;\ldots\vrho_{\rm q}\xnt$.
In these transformations parameters $a'(x;t)$ and $a(x;t)$ are connected by 
the relation
\[
a'(x;t)=a(x;t)-\beta(\bmv{r};t)\lc\frac{p^2}{2m}-
        \bmv{V}(\bmv{r};t)\bmv{p}+\frac m2V^2(\bmv{r};t)\rc.
\]
In the case, when conditions (\ref{e2.18}) take place, one can obtain some
relations, taking into account the self-consistency conditions and
varying the modified Massieu-Planck functional
\[
\Phi(t)=\ln\int\d\mGamma_N\;
        \exp\lc-\int\d\bmv{r}\;\beta(\bmv{r};t)\hat{\cal E}'(\bmv{r})-
        \int\d x\;a'(x;t)\hat{n}_1(x)\rc
\]
with respect to parameters $\beta(\bmv{r};t)$ and $a'(x;t)$
%
%
\be
\begin{array}{lcl}
\ds\frac{\delta\Phi(t)}{\delta\beta(\bmv{r};t)}&=&
        -\la\hat{\cal E}'(\bmv{r})\ra^t_{\rm q}=
        -\la\hat{\cal E}'(\bmv{r})\ra^t,\\
\ds\frac{\delta\Phi(t)}{\delta a(x;t)}&=&
        -\la\hat{n}_1(x)\ra^t_{\rm q}=-\la\hat{n}_1(x)\ra^t=-f_1(x;t).\\
\end{array}
\label{e2.19}
\ee
It means that the parameter $\beta(\bmv{r};t)$ is conjugated to the
average energy in an accompanying reference frame, $a'(x;t)$ is conjugated
to the nonequilibrium oneparticle distribution function $f_1(x;t)$. To
determine the physical meaning of these parameters let us define entropy of
the system taking into account the self-consistency conditions
(\ref{e2.18})
%
%
\be
S(t)=-\la\ln\vrho_{\rm q}\xnt\ra^t_{\rm q}=\Phi(t)+\int\d\bmv{r}\;
        \beta(\bmv{r};t)\la\hat{\cal E}'(\bmv{r})\ra^t+
        \int\d x\;a'(x;t)\la\hat{n}_1(x)\ra^t.\label{e2.20}
\ee
Taking functional derivatives of $S(t)$ (\ref{e2.20}) with respect to
$\la\hat{\cal E}'(\bmv{r})\ra^t$ and $\la\hat{n}_1(x)\ra^t$ at fixed
corresponding averaged values gives the following thermodynamic relations:
%
%
\be
\frac{\delta S(t)}{\delta\la\hat{\cal E}'(\bmv{r})\ra^t}=\beta(\bmv{r};t),
        \quad\frac{\delta S(t)}{\delta f_1(x;t)}=a'(x;t).\label{e2.21}
\ee
Hence, $\beta(\bmv{r};t)$ is an analogue of local inverse temperature.

In general case, when kinetic and hydrodynamic processes are considered
simultaneously, the quasi\-equilibrium distribution function (\ref{e2.16}) 
or (\ref{e2.13}) can be rewritten in somewhat other form. This form is more
convenient
for comparison with $\vrho_{\rm q}\xnt$ (\ref{e2.8}), obtained
in the usual way \cite{4}, when $f_1(x;t)$ is only the 
parameter of the abbreviated description. First of all, let us note that one
can include the parameter $\Phi(t)$ in (\ref{e2.13}) into parameter
$a(x;t)$ as a term, which does not depend on $x$. Parameter $a(x;t)$ in
$\vrho_{\rm q}\xnt$ can be excluded with the help of the self-consistency
condition $\la\hat{n}(x)\ra^t_{\rm q}=\la\hat{n}(x)\ra^t=f_1(x;t)$. Reducing
of $\vrho_{\rm q}\xnt$ results in
%
%
\be
\vrho_{\rm q}\xnt=\exp\lc-U_N(\bmv{r}^N;t)\rc\prod_{l=1}^N
        \frac{f_1(x_l;t)}{u(\bmv{r}_l;t)},\label{e2.26}
\ee
where functions $u(\bmv{r}_l;t)$ are obtained from the relations
%
%
\bea
u(\bmv{r}_l;t)&=&\int\frac{\d\bmv{r}^{N-1}}{(N-1)!}
        \exp\lc-U_N(\bmv{r},\bmv{r}^{N-1};t)\rc
        \prod_{l=2}^N\frac{n(\bmv{r}_l;t)}{u(\bmv{r}_l;t)},
        \label{e2.27}\\
U_N(\bmv{r}^N;t)&=&U_N(\bmv{r}_1,\ldots,\bmv{r}_N;t)=
        \frac 12 \sum_{l\ne j=1}^N\mPhi(|\bmv{r}_l-\bmv{r}_j|)
        \beta(\bmv{r}_j;t).\nonumber,
\eea
$n(\bmv{r};t)=\la\hat{n}(\bmv{r})\ra^t=\int\d\bmv{p}\;f_1(x;t)$ is the
nonequilibrium particles concentration.
In last expression (\ref{e2.26}), $U_N(\bmv{r};t)$ and $u(\bmv{r}_l;t)$
depend explicitly and implicitly, respectively, on $n(\bmv{r};t)$ and
$\beta(\bmv{r};t) $(or $\la\hat{\cal E}'(\bmv{r})\ra^t$). To obtain the
ordinary Bogolubov scheme \cite{4}, it is necessary to put
$U_N(\bmv{r};t)=0$ in (\ref{e2.26}) and (\ref{e2.27}). Then, one can define
$u=\e$, and (\ref{e2.26}) transforms into the quasiequilibrium distribution
(\ref{e2.8}) as it should be. In general case $u(\bmv{r};t)$ is a
functional of nonequilibrium density of particles number
$\la\hat{n}(\bmv{r})\ra^t$ and $\beta(\bmv{r};t)$, which is an analogue of
local inverse temperature. Nevertheless, one should handle with care with
this analogy as far as definition (\ref{e2.26}) can describe states which
are far from local equilibrium. In particular, $f_1(x;t)$ can distinguish
considerably from the local Maxwellian distribution.

The entropy expression (\ref{e2.20}) can be transformed accordingly to the
structure of the quasiequilibrium distribution function (\ref{e2.26})
%
%
\be
S(t)=\int\d\bmv{r}\;\beta(\bmv{r};t)\la\hat{\cal E}_{\rm int}(\bmv{r})\ra^t-
        \int\d x\;f_1(x;t)\ln\frac{f_1(x;t)}{u(\bmv{r};t)}.\label{e2.28}
\ee
Here ``potential'' and ``kinetic'' parts are separated. In the case of
low density gases, the influence of the potential energy can be
neglected and $u(\bmv{r};t)=\e$. Then the expression (\ref{e2.28}) tends
to the usual Boltzmann entropy.

\section{The modified group expansions}
\setcounter{equation}{0}

Integrating equation (\ref{mle}) over
the phase space of $(N-s)$ particles, we obtain the equation chain for the
$s$-particle nonequilibrium distribution function
$f_s\lp x^s;t\rp=\int\d\mGamma_{N-s}\vrho\lp x^N;t\rp$ \cite{1,21}:
%
%
\be
\lp\ddt+\im L_s\rp f_s\lp x^s;t\rp+
        \sum_{j=1}^s\int\d x_{s+1}\;\im L(j,s+1)
        f_{s+1}\lp x^{s+1};t\rp=
        \label{mbec}
\ee
\[
\qquad-\veps\lp f_s\lp x^s;t\rp-g_s\lp\bmv{r}^s;t\rp
        \prod_{j=1}^s f_1\lp x_j;t\rp\rp,
\]
%
%
where $g_s\lp\bmv{r}^s;t\rp=\int\d\mGamma_{N-s}\;\d\bmv{p}^s\;
\vrho_{\rm q}\lp x^N,t\rp$ is the quasiequilibrium $s$-particle coordinate
distribution function, which depends on $n(\bmv{r};t)$ and
$\beta(\bmv{r};t)$ functionally. Due to the fact, that $g_1(\bmv{r}_1;t)=1$,
the equation chain (\ref{mbec}) is distinguished from the ordinary BBGKY
hierarchy \cite{17} by the existence of sources in the right-hand parts
of the equations beginning from the second one. Such sources take into
account both one-particle and collective hydrodynamical effects.

Let us note, that equation chain \refp{mbec} should be amplified by 
equations for spatial quasiequilibrium distribution functions 
$g_s(\bmv{r}^s;t)$, which functionally depend on nonequilibrium density of 
particles number $n(\bmv{r};t)$ and ``inversed'' local temperature 
$\beta(\bmv{r};t)$. Specifically, it has been shown in \cite{x}, that pair 
quasiequilibrium distribution function $g_2(\bmv{r}_1,\bmv{r}_2;t)$ is 
connected with a pair quasiequilibrium correlation function: 
$h_2(\bmv{r}_1,\bmv{r}_2;t)=g_2(\bmv{r}_1,\bmv{r}_2;t)-1$. On its turn, 
$h_2(\bmv{r}_1,\bmv{r}_2;t)$ satisfies the Ornstein-Zernike equation
\[
h_2(\bmv{r}_1,\bmv{r}_2;t)=c_2(\bmv{r}_1,\bmv{r}_2;t)+\int\d\bmv{r}_3\;
	c_2(\bmv{r}_1,\bmv{r}_3;t)h_2(\bmv{r}_2,\bmv{r}_3;t)
	n(\bmv{r}_3;t),
\]
where $c_2(\bmv{r}_1,\bmv{r}_2;t)$ is a direct quasiequilibrium correlation 
function.

To analyze the BBGKY hierarchy, we follow similarly to 
the papers by Zubarev and Novikov \cite{4,18,19,20} and earlier papers
by \cite{5} and Cohen \cite{3,6}, and cross from 
nonequilibrium distribution functions $f_s(x^s;t)$ to irreducible
distribution ones $G_s(x^s;t)$, which may be introduced by equalities 
presented in \cite{4,5}. In our case with some modifications we obtain
%
%
\bea
\lefteqn{\ds f_1(x_1;t)=G_1(x_1;t),}\label{e4.1}\\
\lefteqn{\ds f_2(x_1,x_2;t)=G_2(x_1,x_2;t)+
        g_2(\bmv{r}_1,\bmv{r}_2;t)G_1(x_1;t)G_1(x_2;t),}\nonumber\\
\lefteqn{\ds f_3(x_1,x_2,x_3;t)=G_3(x_1,x_2,x_3;t)+
        \sum_{P}G_2(x_1,x_2;t)G_1(x_3;t)+{}}\nonumber\\
&&g_3(\bmv{r}_1,\bmv{r}_2,\bmv{r}_3;t)G_1(x_1;t)G_1(x_2;t)G_1(x_3;t),
        \nonumber\\
\lefteqn{\ds\vdots}&&\vdots\nonumber
\eea
Here, the position-dependent quasiequilibrium distribution functions
$g_2(\bmv{r}_1,\bmv{r}_2;t),g_3(\bmv{r}_1,\bmv{r}_2,\bmv{r}_3;t)$,
$g_s(\bmv{r}^s;t)$ are defined by relations of paper \cite{1}.
The modification of group expansions (\ref{e4.1}) consists in that a
considerable part of the space time correlations is accumulated in
quasiequilibrium functions $g_s(\bmv{r}^s;t)$. If all $g_s(\bmv{r}^s;t)=1$
for $s=2,3,\ldots$ these group expansions coincide with those of paper
\cite{4}. As far as each line in (\ref{e4.1}) brings exactly for new
functions $G_s(\bmv{r}^s;t)$, $s=1,2,3,\ldots$, the corresponding
equations can be solved with respect to irreducible distribution functions
and we may write the following:
%
%
\bea
\lefteqn{\ds G_1(x_1;t)=f_1(x_1;t),}\label{e4.2}\\
\lefteqn{\ds G_2(x_1,x_2;t)=f_2(x_1,x_2;t)-
        g_2(\bmv{r}_1,\bmv{r}_2;t)f_1(x_1;t)f_1(x_2;t),}\nonumber\\
\lefteqn{\ds G_3(x_1,x_2,x_3;t)=f_3(x_1,x_2,x_3;t)-
        \sum_{P}f_2(x_1,x_2;t)f_1(x_3;t)-{}}\nonumber\\
&&h_3(\bmv{r}_1,\bmv{r}_2,\bmv{r}_3;t)f_1(x_1;t)f_1(x_2;t)f_1(x_3;t).
        \nonumber\\
\lefteqn{\ds\vdots}&&\vdots\nonumber
\eea
In (\ref{e4.1}) and (\ref{e4.2}) the symbol $\ds\sum_P$ denotes the sum of
all different permutations of coordinates for three and more particles,
%
%
\bea
h_3(\bmv{r}_1,\bmv{r}_2,\bmv{r}_3;t)&=&
        g_3(\bmv{r}_1,\bmv{r}_2,\bmv{r}_3;t)-g_2(\bmv{r}_1,\bmv{r}_2;t)-
        g_2(\bmv{r}_1,\bmv{r}_3;t)-g_2(\bmv{r}_2,\bmv{r}_3;t)\label{e4.3}\\
&\equiv&h_3^{\prime}(\bmv{r}_1,\bmv{r}_2,\bmv{r}_3;t)-2,\nonumber
\eea
where $h_3^{\prime}(\bmv{r}_1,\bmv{r}_2,\bmv{r}_3;t)$ is the 
three-particle quasiequilibrium correlation function. Now let us 
write the BBGKY hierarchy \cite{1,2} for irreducible distribution functions 
$G_s(x^s;t)$, namely, the first two equations, 
%
%
\bea
\lefteqn{\ds \lp\ddt+\im L(1)\rp G_1(x_1;t)+{}}\label{e4.4}\\
&&\int\d x_2\;\im L(1,2)g_2(\bmv{r}_1,\bmv{r}_2;t)G_1(x_1;t)G_1(x_2;t)+
        \int\d x_2\;\im L(1,2)G_2(x_1,x_2;t)=0.\nonumber
\eea
Differentiating the relation for $G_2(x_1,x_2;t)$ in (\ref{e4.2}) with
respect to time and using the second equation from the BBGKY hierarchy for
the function $f_2(x_1,x_2;t)$, we can get for the pair irreducible
distribution function $G_2(x_1,x_2;t)$ the equation, which reads
%
%
\bea
\lefteqn{\ds\lp\ddt+\im L_2+\veps\rp G_2(x_1,x_2;t)=
        -\lp\ddt+\im L_2\rp g_2(\bmv{r}_1,\bmv{r}_2;t)G_1(x_1;t)G_1(x_2;t)-{}
        }\label{e4.5}\\
&&\int\d x_3\;\Big\{\im L(1,3)+\im L(2,3)\Big\}\;\Big\{G_3(x_1,x_2,x_3;t)+
        \sum_PG_2(x_1,x_2;t)G_1(x_3;t)+{}\nonumber\\
&&g_3(\bmv{r}_1,\bmv{r}_2,\bmv{r}_3;t)G_1(x_1;t)G_1(x_2;t)G_1(x_3;t)\Big\}.
        \nonumber
\eea
In a similar way, we can obtain other equations for the three-particle
irreducible function $G_3(x_1,x_2,x_3;t)$ and higher $G_s(x^s;t)$ ones.
One remembers now, that appearing the quasiequilibrium distribution
functions $g_2(\bmv{r}_1,\bmv{r}_2;t)$,
$g_3(\bmv{r}_1,\bmv{r}_2,\bmv{r}_3;t)$, $g_s(\bmv{r}^s;t)$ in the hierarchy
is closely connected with the fact that the boundary conditions for
solutions of the Liouville equation take into consideration both the
nonequilibrium character of one-particle distribution function and local
conservation laws, that corresponds to a consistent description of
kinetics and hydrodynamics of the system \cite{1,21}. Since in the present
paper we analyze two first equations (\ref{e4.4}) and (\ref{e4.5}) only, we
will not write down next ones. It is important to note, that if we put
formally $g_s(\bmv{r}^s;t)\equiv 1$ for all $s=2,3,\ldots$ in (\ref{e4.4})
and (\ref{e4.5}), we come to the first two equations of the BBGKY
hierarchy for irreducible distribution functions $G_1(x_1;t)$ and
$G_2(x_1,x_2;t)$, which were obtained in the paper \cite{4} by D.N.Zubarev
and M.Yu.Novikov. The first term in the right-hand side of (\ref{e4.5}) is a
peculiarity of (\ref{e4.4}) and (\ref{e4.5}) equations system. This is the
term with time derivative of the pair quasiequilibrium distribution
function $g_2(\bmv{r}_1,\bmv{r}_2;t)$. As it was shown in \cite{1,21,22},
the pair quasiequilibrium distribution function is a functional of local
values of the temperature $\beta(\bmv{r};t)$ and the mean particle density
$n(\bmv{r};t)$. Thus, time derivatives of
$g_2\big(\bmv{r}_1,\bmv{r}_2|\beta(t),n(t)\big)$ will be conformed to
$\beta(\bmv{r};t)$ and $n(\bmv{r};t)$. These quantities, in its turn,
according to the self-consistency conditions \cite{22}, will be expressed
via the average energy value $\La\hat{\cal E}'(\bmv{r})\Ra^t$ in an
accompanying reference frame and via $\la \hat{n}(\bmv{r})\ra^t$, which
constitute a basis of the hydrodynamical description of nonequilibrium state
of the system.

If we neglect the term in the right-hand side of equation (\ref{e4.5}),
which takes into account ternary correlations between particles, we shall
obtain from (\ref{e4.5}) the equation in the so-called ``pair''collisions
approximation with the following form:
%
%
\be
\lp\ddt+\im L_2+\veps\rp G_2(x_1,x_2;t)=-\lp\ddt+
        \im L_2\rp g_2(\bmv{r}_1,\bmv{r}_2;t)G_1(x_1;t)G_1(x_2;t).
        \label{e4.6}
\ee
The formal solution to this equation reads
%
%
\be
G_2(x_1,x_2;t)=-\intl_{-\infty}^t\d t'\;\e^{\ds(\veps+\im L_2)(t-t')}
        \ls\dd{t'}+\im L_2\rs g_2(\bmv{r}_1,\bmv{r}_2;t')
        G_1(x_1;t')G_1(x_2;t').
        \label{e4.7}
\ee
Inserting this solution into the first equation (\ref{e4.4}) we obtain the
following kinetic equation for one-particle distribution function
$f_1(x_1;t)=G_1(x_1;t)$:
%
%
\bea
\lefteqn{\ds \lp\ddt+\im L(1)\rp f_1(x_1;t)=-\int\d x_2\;\im L(1,2)
        g_2(\bmv{r}_1,\bmv{r}_2;t)f_1(x_1;t)f_1(x_2;t)}\label{e4.8}\\
&&{}-\int\d x_2\;\im L(1,2)\intl_{-\infty}^{t}\d t'\;
        \e^{\ds(\veps+\im L_2)(t-t')}
        \ls\dd{t'}+\im L_2\rs g_2(\bmv{r}_1,\bmv{r}_2;t')
        f_1(x_1;t')f_1(x_2;t'),\nonumber
\eea
where the first term in the right-hand side is a generalization of the
Vlasov mean field and corresponds to kinetic mean field theory -- KMFT 
\cite{23,24}. The kinetic equation (\ref{e4.8}) is completely equivalent to 
that of \cite{1}, which is obtained in the ``pair'' collisions approximation.

Let us consider now the set of equations (\ref{e4.4}) and (\ref{e4.5}) in
such an approximation that the three-particle
irreducible distribution function $G_3(x_1,x_2,x_3;t)$ and 
$h_3(\bmv{r}_1,\bmv{r}_2,\bmv{r}_3;t)$ are neglected
in the second equation (\ref{e4.5}). This is in the spirit of
the polarization approximation introduced for obtaining the Lenard-Balescu 
kinetic equation for a homogeneous Coulomb plasma \cite{14}. Taking into 
account (\ref{e4.3}) as well as the relation
$G_3(x_1,x_2,x_3;t)\equiv 0$ and similarly
$h_3(\bmv{r}_1,\bmv{r}_2,\bmv{r}_3;t)\equiv 0$, we rewrite the equation
(\ref{e4.5}) in the form:
%
%
\bea
\lefteqn{\ds \lp\ddt+\im L_2+\veps\rp G_2(x_1,x_2;t)=
        -\lp\ddt+\im L_2\rp
        g_2(\bmv{r}_1,\bmv{r}_2;t)G_1(x_1;t)G_1(x_2;t)}\label{e4.9}\\
\lefteqn{\ds{}-\int\d x_3\;\im L(1,3)\Big\{G_2(x_1,x_2;t)G_1(x_3;t)+
        G_2(x_2,x_3;t)G_1(x_1;t)\Big\}}\nonumber\\
\lefteqn{\ds{}-\int\d x_3\;\im L(2,3)\Big\{G_2(x_1,x_2;t)G_1(x_3;t)+
        G_2(x_1,x_3;t)G_1(x_2;t)\Big\}}\nonumber\\
\lefteqn{\ds{}-\int\d x_3\Big\{\im L(1,3)+\im L(2,3)\Big\}
        \Big\{g_2(\bmv{r}_1,\bmv{r}_2;t)+g_2(\bmv{r}_1,\bmv{r}_3;t)+
        g_2(\bmv{r}_2,\bmv{r}_3;t)\Big\}G_1(x_1;t)G_1(x_2;t)G_1(x_3;t).}
        \nonumber
\eea
Next, let us introduce the operator, which can be obtained by variation of
the Vlasov collision integral near a nonequilibrium distribution 
$G_1(x_1;t)$:
%
%
\bea
\delta\lp\int\d x_3\;\im L(1,3)G_1(x_3;t)G_1(x_1;t)\rp&=&\label{e4.10}\\
\int\d x_3\;\im L(1,3)G_1(x_3;t)\delta G_1(x_1;t)\,&=&
        \El(x_1;t)\delta G_1(x_1;t).\nonumber
\eea
Then one represents the equation \refp{e4.9} with the help of operator
$\El(x_1;t)$ in the form
%
%
\bea
\lefteqn{\ds\lp\ddt+\im L_2+\El(x_1,x_2;t)+\veps\rp G_2(x_1,x_2;t)={}}
        \label{e4.11}\\
&&{}-\lp\ddt+\im L_2+\El(x_1,x_2;t)\rp g_2(\bmv{r}_1,\bmv{r}_2;t)
        G_1(x_1;t)G_1(x_2;t),\nonumber
\eea
from which the formal solution for the irreducible two-particle distribution
function reads
%
%
\bea
\lefteqn{\ds G_2(x_1,x_2;t)=-\intl_{-\infty}^{t}\d t'\;\e^{\veps(t'-t)}
        U(t,t')\times{}}\label{e4.12}\\
&&\lp\dd{t'}+\im L_2+\El(x_1,x_2;t')\rp g_2(\bmv{r}_1,\bmv{r}_2,t')
        G_1(x_1;t')G_1(x_2;t'),\nonumber
\eea
and $U(t,t')$ is the evolution operator,
%
%
\be
U(t,t')=\exp_+\lc-\intl_{t'}^{t}\d t''\;\lp\im L_2+\El(x_1,x_2;t'')\rp\rc,
        \qquad\El(x_1,x_2;t)=\El(x_1;t)+\El(x_2;t).\label{e4.13}
\ee
As a result, we obtain the expression for the irreducible quasiequilibrium
two-particle distribution function $G_2(x_1,x_2;t)$ in the generalized 
polarization approximation. Inserting this expression \refp{e4.12} into the 
first equation of the chain \refp{e4.4} yields:
%
%
\bea
\lefteqn{\ds\lp\ddt+\im L(1)\rp G_1(x_1;t)+\int\d x_2\;\im L(1,2)
        g_2(\bmv{r}_1,\bmv{r}_2;t)G_1(x_1;t)G_1(x_2;t)={}}\label{e4.14}\\
&&\int\d x_2\!\intl_{-\infty}^{t}\d t'\;\e^{\veps(t'-t)}\im L(1,2)U(t,t')
        \lp\dd{t'}+\im L_2+\El(x_1,x_2;t')\rp g_2(\bmv{r}_1,\bmv{r}_2;t')
        G_1(x_1;t')G_1(x_2;t').\nonumber
\eea
This is the kinetic equation for the nonequilibrium oneparticle distribution
function with the non-Markovian collision integral in the generalized
polarization approximation. It is necessary to note, that the presence of
the Vlasov's operator $\El(x_1,x_2;t)$ in the collision integral
\refp{e4.14} indicates about taking into consideration collective effects.
An analysis of the collision integral \refp{e4.14} in the general case is 
rather a complicated problem. But it is obvious, that the collision integral
in \refp{e4.14}, or the expression for $G_2(x_1,x_2;t)$ in \refp{e4.12} may
be much simplified for every physical model of a particle system, or for
each nonequilibrium state of the collision integral in \refp{e4.14}. To show
this, we shall consider two concrete cases: a hard spheres model and a
Coulomb plasma.

\section{Hard spheres models in the polarization approximation}
\setcounter{equation}{0}

In this section we shall perform the investigation of kinetic processes for
hard spheres model in approximations, which are higher than ``pair''
collisions one. We take into account the character of model parameters
and the results of the previous section of this article and papers 
\cite{13,14,25}. This investigation is convenient to carry out on the basis
of equation chain \refp{e4.4}, \refp{e4.5} at formal substitution
of a potential part of the Liouville operator $\im L(1,2)$ \cite{1,2} by the 
Enskog collision operator $\hat{T}(1,2)$ \cite{1,2}. In this case, equations
\refp{e4.4} and \refp{e4.5} have the form
%
%
\bea
\lefteqn{\ds \lp\ddt+\im L(1)\rp G_1(x_1;t)+{}}\label{e4.15}\\
&&\int\d x_2\;\hat{T}(1,2)g_2(\bmv{r}_1,\bmv{r}_2;t)G_1(x_1;t)G_1(x_2;t)+
        \int\d x_2\;\hat{T}(1,2)G_2(x_1,x_2;t)=0,\nonumber
\eea
%
%
%
%
%
\bea
\lefteqn{\ds\lp\ddt+\im L_2^0+\hat{T}(1,2)+
        \veps\rp G_2(x_1,x_2;t)={}}\label{e4.16}\\
&&{}-\lp\ddt+\im L_2^0+\hat{T}(1,2)\rp g_2(\bmv{r}_1,\bmv{r}_2;t)
        G_1(x_1;t)G_1(x_2;t)-\int\d x_3\;\Big\{\hat{T}(1,3)+
        \hat{T}(2,3)\Big\}\nonumber\\
&&{}\times\Big\{G_3(x_1,x_2,x_3;t)+\sum_PG_2(x_1,x_2;t)G_1(x_3;t)+
        g_3(\bmv{r}_1,\bmv{r}_2,\bmv{r}_3;t)G_1(x_1;t)
        G_1(x_2;t)G_1(x_3;t)\Big\}.\nonumber
\eea
Further, we will consider the same approximations concerning to equation
\refp{e4.16}, in which $G_3(x_1,x_2,x_3;t)$ and
$h_3(\bmv{r}_1,\bmv{r}_2,\bmv{r}_3;t)$ are neglected. Then, if we introduce
similarly to \refp{e4.10} the Boltzmann-Enskog collision operator
$C(x_1;t)$, the equation \refp{e4.16} could be rewritten in the next form:
%
%
\be
\delta\int\d x_3\;\hat{T}(1,3)G_1(x_1;t)G_1(x_3;t)=
        C(x_1;t)\delta G_1(x_1;t),\label{e4.17}
\ee
%
%
%
%
%
\bea
\lefteqn{\ds\lp\ddt+\im L_2^0+\hat{T}(1,2)+C(x_1,x_2;t)+
        \veps\rp G_2(x_1,x_2;t)={}}
        \label{e4.18}\\
&&{}-\lp\ddt+\im L_2^0+\hat{T}(1,2)+C(x_1,x_2;t)\rp
        g_2(\bmv{r}_1,\bmv{r}_2;t)G_1(x_1;t)G_1(x_2;t),\nonumber
\eea
Hence it appears, that the formal solution to $G_2(x_1,x_2;t)$ reads
%
%
\bea
\lefteqn{\ds G_2(x_1,x_2;t)=-\intl_{-\infty}^0\d t'\;\e^{\veps(t'-t)}
        U_{\rm hs}(t,t')\times{}}\label{e4.19}\\
&&\lc\dd{t'}+\im L_2^0+\hat{T}(1,2)+C(x_1,x_2;t')\rc
        g_2(\bmv{r}_1,\bmv{r}_2;t')G_1(x_1;t')G_1(x_2;t'),\nonumber
\eea
where $U_{\rm hs}(t,t')$ is the evolution operator for the system of hard
spheres:
%
%
\be
U_{\rm hs}(t,t')=\exp_+\lc-\intl_{t'}^{t}\d t''\;
        \ls\im L_2^0+\hat{T}(1,2)+C(x_1,x_2;t'')\rs\rc,\label{e4.20}
\ee
\[
C(x_1,x_2;t)=C(x_1;t)+C(x_2;t).
\]
Now let us put \refp{e4.19} into the first equation \refp{e4.15}. Then the
resulting equation looks:
%
%
\bea
\lefteqn{\ds\lp\ddt+\im L(1)\rp G_1(x_1;t)=\int\d x_2\;\hat{T}(1,2)
        g_2(\bmv{r}_1,\bmv{r}_2;t)G_1(x_1;t)G_1(x_2;t)-
        \int\d x_2\;\hat{T}(1,2)\times{}}\label{e4.21}\\
&&\ds\intl_{-\infty}^0\d t'\;\e^{\veps(t'-t)}U_{\rm hs}(t,t')
        \lc\dd{t'}+\im L_2^0+\hat{T}(1,2)+C(x_1,x_2;t')\rc
        g_2(\bmv{r}_1,\bmv{r}_2;t')G_1(x_1;t')G_1(x_2;t').\nonumber
\eea
This equation can be called as a generalized kinetic equation for the
nonequilibrium one-particle distribution function of hard spheres with the
non-Markovian collision integral in the generalized polarization
approximation. The first term in the right-hand side of this equation is the
collision integral from the revised Enskog theory \cite{26}. Neglecting time
retardation effects and assuming that the operator $C(x_1,x_2;t)$ does not
depend on time when
\[G_1(x_1;t)=f_0(\bmv{p})=\lp\frac{m}{2\pi
kT}\rp^{3/2}\exp\lc-\frac{p^2}{2mkT}\rc
\]
is the local equilibrium Maxwell distribution function, the next term can
be rewritten in a simplified form,
%
%
\be
I_{\rm R}(x_1;t)=-\intl_{-\infty}^{t}\d t'\;
        \e^{\veps(t'-t)}R_0(x_1;t,t')G_1(x_1;t')
        -\intl_{-\infty}^{t}\d t'\;\e^{\veps(t'-t)}R_1(x_1;t,t')G_1(x_1;t'),
        \label{e4.22}
\ee
where
%
%
\be
R_0(x_1;t,t')\!=\!\int\d x_2\,\hat{T}(1,2)\e^{\lc(t'-t)
        \lp\im L_2^0+\hat{T}(1,2)+C(x_1,x_2)\rp\rc}\!\!
        \ls\im L_2^0+\!C(x_1,x_2)\rs\! 
        g_2(\bmv{r}_1,\bmv{r}_2;t')G_1(x_2;t'),
        \label{e4.23}
\ee
\be
R_1(x_1;t,t')\!=\!\int\d x_2\,\hat{T}(1,2)\e^{\lc(t'-t)
        \lp\im L_2^0+\hat{T}(1,2)+C(x_1,x_2)\rp\rc}
        \hat{T}(1,2)g_2(\bmv{r}_1,\bmv{r}_2;t')G_1(x_2;t'),\label{e4.24}
\ee
$R_1(x_1;t,t')$ is the generalized ring operator. The kinetic equation
\refp{e4.21} together with \refp{e4.23}, \refp{e4.24} is the generalization
of the kinetic equation for a system of hard spheres, which has been
obtained by Bogolubov in \cite{14,25}. It coincides with that, when
the quasiequilibrium pair distribution function of hard spheres is set
formally to be unity.

\section{Coulomb plasma in the polarization approximation}
\setcounter{equation}{0}

Here we shall study an electron gas, which is contained into a homogeneous
positively charged equilibrating background. This background can be created,
for example, by hard motionless ions. Then, electrons interact
according to the Coulomb law:
\[
\mPhi(|\bmv{r}_{12}|)=\frac{e^2}{|\bmv{r}_1-\bmv{r}_2|}=
        \frac{e^2}{|\bmv{r}_{12}|},
\]
the Fourier transform of which exists in the form of a real function
$\mPhi(|\bmv{k}|)$:
%
%
\be
\frac{e^2}{r_{12}}=\int\frac{\d\bmv{k}}{(2\pi)^3}\frac{4\pi e^2}{k^2}
        \e^{\im\bmv{k}\cdot\bmv{r}_{12}},\quad\mPhi(k)=\frac{4\pi e^2}{k^2},
        \label{e4.25}
\ee
here $\bmv{k}$ is a wavevector, $e$ is the electron charge. Let us
consider equation chain \refp{e4.4}, \refp{e4.9} in the homogeneous case,
when $G_1(x_1;t)=G_1(\bmv{p}_1;t)$ and pair distribution functions 
depend on $|\bmv{r}_{12}|$. Following the Bogolubov method \cite{14,17}, we
shall suppose that the one-particle distribution function $G_1(\bmv{p}_1;t)$
is calculated in the ``zeroth'' order on interaction constant $q$, pair
distribution functions $G_2(\bmv{r}_{12},\bmv{p}_1,\bmv{p}_2;t)$ and
$g_2(\bmv{r}_{12};t)$ as the first order on $q$, and $G_3(x_1,x_2,x_3;t)$,
$g_3(\bmv{r}_1,\bmv{r}_2,\bmv{r}_3;t)$ $\sim$ $q^2$,
where $q=\frac{e^2}{r_{\rm d}}\mTheta$, $r_{\rm d}=\sqrt{\mTheta/4\pi e^2n}$
is the Debye radius, $n=N/V$, $\mTheta=k_{\rm B}T$, $k_{\rm B}$ is the
Boltzmann constant, $T$ is thermodynamic 
temperature. Therefore, to obtain an equation
for $G_2(\bmv{r}_{12},\bmv{p}_1,\bmv{p}_2;t)$ in the first approximation on
interaction constant $q$ without time retardment effects, it is necessary
to retain all integral terms, but omit all the others. In this case,
using the Fourier transform with respect to spatial coordinates for a
homogeneous Coulomb electron gas, the set of equations \refp{e4.4},
\refp{e4.9} yields:
\[
\ddt G_1(\bmv{p}_1;t)=-\dd{\bmv{p}_1}\!\int\!\d\bmv{k}\d\bmv{p}_2
        \im\mPhi(|\bmv{k}|)g_2(\bmv{k};t)
        G_1(\bmv{p}_1;t)G_1(\bmv{p}_2;t)\!-\!
        \dd{\bmv{p}_1}\!\int\!\d\bmv{k}\d\bmv{p}_2
        \im\mPhi(|\bmv{k}|)G_2(\bmv{k},\bmv{p}_1,\bmv{p}_2;t),
\]
or
%
%
\be
\ddt G_1(\bmv{p}_1;t)=\dd{\bmv{p}_1}G_1(\bmv{p}_1;t)\int\d\bmv{k}\;\bmv{k}
        \mPhi(|\bmv{k}|)\Imm g_2(\bmv{k};t)+
        \dd{\bmv{p}_1}\int\d\bmv{k}\;\bmv{k}\mPhi(|\bmv{k}|)\Imm
        G_2(\bmv{k},\bmv{p}_1;t)\label{e4.26}
\ee
and equation for $G_2(\bmv{k},\bmv{p}_1,\bmv{p}_2;t)$:
%
%
\bea
\lefteqn{\ds\lp\ddt+\im\bmv{k}\frac{\bmv{p}_{12}}{m}+\veps\rp
        G_2(\bmv{k},\bmv{p}_1,\bmv{p}_2;t)={}}\label{e4.27}\\
&&\im\bmv{k}\mPhi(|\bmv{k}|)\lc\dd{\bmv{p}_1}G_1(\bmv{p}_1;t)
        \int\d\bmv{p}_3\;G_2(\bmv{k},\bmv{p}_2,\bmv{p}_3;t)-
        \dd{\bmv{p}_2}G_1(\bmv{p}_2;t)\int\d\bmv{p}_3\;
        G_2(\bmv{k},\bmv{p}_1,\bmv{p}_3;t)\rc+{}\nonumber\\
&&\im\bmv{k}\mPhi(|\bmv{k}|)\lc\dd{\bmv{p}_1}G_1(\bmv{p}_1;t)
        g_2(-\bmv{k};t)G_1(\bmv{p}_2;t)-
        \dd{\bmv{p}_2}G_1(\bmv{p}_2;t)g_2(\bmv{k};t)G_1(\bmv{p}_1;t)\rc,
        \nonumber
\eea
$\veps\to+0$, and $G_2(\bmv{k},\bmv{p}_1;t)=
\int\d\bmv{p}_2\;G_2(\bmv{k},\bmv{p}_1,\bmv{p}_2;t)$; $\Imm
g_2(\bmv{k};t)$, $\Imm G_2(x_1,x_2;t)$ are imaginary parts of the
corresponding distribution functions. The following properties should be
noted:
\bean
G_2(-\bmv{k},\bmv{p}_1,\bmv{p}_2;t)&=&G_2^*(\bmv{k},\bmv{p}_1,\bmv{p}_2;t),\\
g_2(-\bmv{k};t)&=&g_2^*(\bmv{k};t).
\eean
The solution to \refp{e4.27}, neglecting time retardment effects, reads:
%
%
\bea
G_2(\bmv{k},\bmv{p}_1,\bmv{p}_2;t)=&&\!\!\!\!\!\!\!\!\!\!\!
        \frac{\bmv{k}\mPhi(|\bmv{k}|)}{\ds\bmv{k}\cdot\frac{\bmv{p}_{12}}{m}-
        \im 0}\lc\dd{\bmv{p}_1}G_1(\bmv{p}_1;t)
        G_2(-\bmv{k},\bmv{p}_2;t)-\dd{\bmv{p}_2}G_1(\bmv{p}_2;t)
        G_2(\bmv{k},\bmv{p}_1;t)\rc+{}\label{e4.28}\\
&&\!\!\!\!\!\!\!\!\!\!\!
        \frac{\bmv{k}\mPhi(|\bmv{k}|)}{\ds\bmv{k}\cdot\frac{\bmv{p}_{12}}{m}-
        \im 0}\lc\dd{\bmv{p}_1}G_1(\bmv{p}_1;t)g_2(-\bmv{k};t)
        G_1(\bmv{p}_2;t)-\dd{\bmv{p}_2}G_1(\bmv{p}_2;t)g_2(\bmv{k};t)
        G_1(\bmv{p}_1;t)\rc.\nonumber
\eea
It should be noticed also that equation \refp{e4.26} contains an
imaginary part of the pair irreducible nonequilibrium distribution function,
to be integrated with respect to momentum of the second particle. Now one
integrates equation \refp{e4.28} over all values of momentum $\bmv{p}_2$
and defines in such a way some function $G_2(\bmv{k},\bmv{p}_1;t)$:
%
%
\bea
\lefteqn{\ds\lp 1+\int\d\bmv{p}_2\;\frac{\bmv{k}\mPhi(|\bmv{k}|)}
        {\ds\bmv{k}\cdot\frac{\bmv{p}_{12}}{m}-\im 0}
        \dd{\bmv{p}_2}G_1(\bmv{p}_2;t)\rp
        G_2(\bmv{k},\bmv{p}_1;t)={}}\label{e4.29}\\
&&\dd{\bmv{p}_1}G_1(\bmv{p}_1;t)\int\d\bmv{p}_2\;
        \frac{\bmv{k}\mPhi(|\bmv{k}|)}
        {\ds\bmv{k}\cdot\frac{\bmv{p}_{12}}{m}-\im 0}
        G_2(-\bmv{k},\bmv{p}_2;t)+{}\nonumber\\
&&\int\d\bmv{p}_2\;\frac{\bmv{k}\mPhi(|\bmv{k}|)}
        {\ds\bmv{k}\cdot\frac{\bmv{p}_{12}}{m}-\im 0}
        \lc\dd{\bmv{p}_1}G_1(\bmv{p}_1;t)g_2(-\bmv{k};t)
        G_1(\bmv{p}_2;t)-\dd{\bmv{p}_2}G_1(\bmv{p}_2;t)g_2(\bmv{k};t)
        G_1(\bmv{p}_1;t)\rc.\nonumber
\eea
Further, we should exclude from \refp{e4.29} the term with
$G_2(-\bmv{k},\bmv{p}_2;t)$. To do this, we follow Lenard \cite{29,30} and
integrate the equation \refp{e4.29} over momentum component
$\bmv{p}_{1\perp}$, which is perpendicular to wavevector $\bmv{k}$.
Resulting expression then reads:
%
%
\bea
\lefteqn{\ds\ls 1+\mPhi(|\bmv{k}|)\Chi(\bmv{k},p_1;t)\rs
        G_2(\bmv{k},p_1;t)=\dd{\bmv{p}_1}G_1(p_1;t)\int\d\bmv{p}_2\;
        \frac{\bmv{k}\mPhi(|\bmv{k}|)}{\ds k\cdot\frac{p_{12}}{m}-\im 0}
        G_2(-\bmv{k},\bmv{p}_2;t)+{}}\label{e4.30}\\
\lefteqn{\ds\int\d\bmv{p}_2\;\frac{\bmv{k}\mPhi(|\bmv{k}|)}
        {\ds k\cdot\frac{p_{12}}{m}-\im 0}
        \lc\dd{\bmv{p}_1}G_1(p_1;t)g_2(-\bmv{k};t)G_1(\bmv{p}_2;t)-
        \dd{\bmv{p}_2}G_1(\bmv{p}_2;t)g_2(\bmv{k};t)G_1(p_1;t)\rc,}
        \nonumber
\eea
Here the following conventional designations have been introduced:
%
%
\bea
\Chi(\bmv{k},p_1;t)&=&\int\d\bmv{p}_2\;\frac{\bmv{k}}
        {\ds k\cdot\frac{p_{12}}{m}-\im 0}\dd{\bmv{p}_2}G_1(\bmv{p}_2;t),
        \quad p_1=\frac{\bmv{p}_1\cdot\bmv{k}}{k},\quad
        p_2=\frac{\bmv{p}_2\cdot\bmv{k}}{k},\quad
        k=|\bmv{k}|,\label{e4.31}\\
G_1(p_1;t)&=&\int\d\bmv{p}_{1\perp}G_1(\bmv{p}_1;t),\nonumber\\
G_2(\bmv{k},p_1;t)&=&\int\d\bmv{p}_{1\perp}\;
        G_2(\bmv{k},\bmv{p}_1;t).\nonumber
\eea
Now we multiply both equations \refp{e4.29} and \refp{e4.30} on
$\dd{p_1}G_1(p_1;t)$ and $\dd{p_1}G_1(\bmv{p}_1;t)$, respectively, and
subtract them:

%
%
\bea
\lefteqn{\ds\,\Big(1+\mPhi(|\bmv{k}|)\Chi(\bmv{k},p_1;t)\Big)
        \ls G_2(\bmv{k},\bmv{p}_1;t)\dd{p_1}G_1(p_1;t)-
        G_2(\bmv{k},p_1;t)\dd{p_1}G_1(\bmv{p}_1;t)\rs={}}\label{e4.32}\\
\lefteqn{\ds\mPhi(|\bmv{k}|)\Chi(\bmv{k},p_1;t)g_2(\bmv{k};t)
        \ls G_1(p_1;t)\dd{p_1}G_1(\bmv{p}_1;t)-
        G_1(\bmv{p}_1;t)\dd{p_1}G_1(p_1;t)\rs.}\nonumber
\eea
If we extract imaginary part of this equation, one can find unknown
quantity $\Imm G_2(\bmv{k},\bmv{p}_1;t)$, provided
$\Imm G_2(\bmv{k},p_1;t)=0$ \cite{29}:
%
%
\bea
\!\!\!\dd{p_1}G_1(p_1;t)\Imm G_2(\bmv{k},\bmv{p}_1;t)=
        \frac{\mPhi(|\bmv{k}|)\Imm
        \ls\Chi(\bmv{k},p_1;t)g_2(\bmv{k};t)\rs}
        {\left|1+\mPhi(|\bmv{k}|)\Chi(\bmv{k},p_1;t)\right|^2}
        \nonumber\\
\!\!\!{}\times\ls G_1(p_1;t)\dd{p_1}G_1(\bmv{p}_1;t)-G_1(\bmv{p}_1;t)
        \dd{p_1}G_1(p_1;t)\rs.\label{e4.33}
\eea
Since $\Imm \Chi(\bmv{k},p_1;t)=-\pi\dd{p_1}G_1(p_1;t)$
\cite{29,30}, putting the expression for $\Imm G_2(\bmv{k},\bmv{p}_1;t)$
into the equation \refp{e4.26} gives the generalized
Bogolubov-Lenard-Balescu kinetic equation for an electron gas in an
equilibrating background
%
%
\bea
\ddt G_1(\bmv{p}_1;t)&=&\dd{\bmv{p}_1}G_1(\bmv{p}_1;t)
        \int\d\bmv{k}\;\bmv{k}\mPhi(|\bmv{k}|)\Imm g_2(\bmv{k};t)+{}
        \label{e4.34}\\
&&\dd{\bmv{p}_1}\int\d\bmv{p}_2\;Q(\bmv{p}_1,\bmv{p}_2;t)
        \ls\dd{\bmv{p}_1}-\dd{\bmv{p}_2}\rs G_1(\bmv{p}_1;t)G_1(\bmv{p}_2;t),
        \nonumber
\eea
where $Q(\bmv{p}_1,\bmv{p}_2;t)$ is a second rank tensor
%
%
\be
Q(\bmv{p}_1,\bmv{p}_2;t)=-\pi\int\d\bmv{k}\;
        \frac{\left|\mPhi(|\bmv{k}|)\right|^2\bmv{k}\cdot\bmv{k}}
        {\left|1+\mPhi(|\bmv{k}|)\Chi(\bmv{k},p_1;t)\right|^2}\Imm
        g_2(\bmv{k};t)\delta\big(\bmv{k}\cdot(\bmv{p}_1-\bmv{p}_2)\big),
        \label{4.35}
\ee
which coincides with $Q(\bmv{p}_1,\bmv{p}_2)$ \cite{28} at
$\Imm g_2(\bmv{k};t)=1$. In this case, the kinetic equation \refp{e4.34}
transforms into the well-known Lenard-Balescu equation \cite{28,29,30}.
Evidently, the generalized Bogolubov-Lenard-Balescu kinetic equation
\refp{e4.34} climes the description of a dense electron gas, since in the
both generalized mean field and generalized Bogolubov-Lenard-Balescu
collision integrals, manyparticle correlations are treated by imaginary
part of $g_2(\bmv{k};t)$. Nevertheless, the problem of divergence in the
collision integral of equation \refp{e4.34} at small distances
($k\to\infty$) still remains. 
There are papers, where divergence of collision integrals is avoided with
the help of special choice of differential cross section (quantum systems
\cite{31}), or via combination of simpler collision integrals (classical
systems \cite{12}). These generalization for collision integrals are
attractive by theirs simplicity and are usable for ideal plasma. But, 
contrary to the obtained by us Bogolubov-Lenard-Balescu kinetic equation, 
they do not work for nonideal plasma. In accordance to the proposed structure
of collision integral
\[
I_{\rm total}=I_{\rm Boltzmann}-I_{\rm Landau}+I_{\rm Lenard-Balescu}
\]
the influence of particles interaction into plasma energy will be defined by
correlation function $g_2(r)$. Its asymptotic is
\[
\lim_{r\to\infty}\ls
        \underbrace{\exp\lc-\frac{e_ae_b}{rk_{\rm B}T}\rc-1}_{\rm Boltzmann}+
        \underbrace{\frac{e_ae_b}{rk_{\rm B}T}}_{\rm Landau}-
        \underbrace{\frac{e_ae_b}{rk_{\rm B}T}
        \e^{-r/r_{\rm D}}}_{\rm Lenard-Balescu}
        \rs=\frac{1}{r^2}\;\;(!),\;\;\mbox{where } r_{_{\rm D}}
        \mbox{ denotes the Debye radius.}
\]
At other combinations one arrives at false expressions for thermodynamic
functions \cite{12}. Dynamical screening, appearing in the obtained by us
generalized Bogolubov-Lenard-Balescu collision integral, is free of these
discrepancies. Generally speaking, the problem of divergency could be solved 
within a frame of charged hard spheres model, combining the results of this 
section and previous one. But this step constitutes an intricate and 
complicated problem and needs a separate consideration.

Evidently, an investigation of the obtained kinetic equation 
is important in view of its solutions and studying transport coefficients 
and time correlation functions for model systems.

\section{Conclusion}
\setcounter{equation}{0}

In view of dense systems study, where the consideration of
spatially interparticle correlations is important, the BBGKY hierarchy
\refp{e4.4}, \refp{e4.5} with the modified boundary conditions and group
expansions has a quite good perspective. The kinetic equation \refp{e4.21}
-- \refp{e4.24} is a generalization of the Bogolubov one \cite{14,25} for a
system of hard spheres. M.Ernst and J.Dorfman \cite{27} had investigated
collective modes in a nonhomogeneous gas and showed, that the solution of a
dispersion equation for hydrodynamic modes leads to the nonanalytic
frequency dependence on wavevector. This is connected with the fact, that
the ring operator for nonhomogeneous systems at small wavenumbers has a
term proportional to $\sqrt{k}$. Similar investigations of collective modes
and time correlation functions in the hydrodynamic region have been carried
out by Bogolubov \cite{14,25,28}. Nevertheless, it is necessary to carry
out analogous investigations of hydrodynamic collective modes and time
correlation functions on the basis of kinetic equation \refp{e4.21}, taking
into account \refp{e4.22} -- \refp{e4.24}, where some part of space
correlations is considered in the pair quasiequilibrium function
$g_2(\bmv{r}_1,\bmv{r}_2;t)$. Obviously, these results may appear to be good
for very dense gases, which could be described by a hard spheres model. An
important factor is that in the kinetic equation \refp{e4.21} --
\refp{e4.24} as well as in the generalized Bogolubov-Lenard-Balescu one,
collective effects are taking into account both via Vlasov's mean field
and pair quasiequilibrium correlation function, which is a functional of
nonequilibrium values of temperature and chemical potential.

Transferring the obtained results on quantum systems is not obvious. Such a
procedure is rather complicated and needs additional investigations. 
Nevertheless, some steps in this way have been done already by our 
colleagues \cite{33,34}.

\end{document}